\documentclass[twocolumn]{aastex63}

\newcommand{\mm}{{\rm \mu m}}
\newcommand{\lsun}{L_\odot}

\newcommand{\msun}{M_\odot}
\newcommand{\msunyr}{M_\odot~{\rm yr}^{-1}}
\usepackage{here}

\usepackage[version=4]{mhchem}
\usepackage{threeparttable}
\usepackage{lipsum}
\usepackage{multirow}
\usepackage{color}

\accepted{April 1 2022}

\submitjournal{ApJ}

\shorttitle{A UIR feature in WR 125}
\shortauthors{Endo et al.}

\graphicspath{{./}{figures/}}

\begin{document}

\title{Detection of a broad 8\,$\mm$ UIR feature in the mid-infrared spectrum of WR 125 observed with Subaru/COMICS}

\correspondingauthor{Izumi Endo}
\email{endo@astron.s.u-tokyo.ac.jp}

\author[0000-0002-9129-5988]{Izumi Endo}
\affiliation{Department of Astronomy, Graduate School of Science, The University of Tokyo, Bunkyo-ku, Tokyo 113-0033, Japan}

\author{Ryan M. Lau}
\affiliation{Institute of Space and Astronautical Science, Japan Aerospace Exploration Agency, Sagamihara City, Kanagawa 252-5210, Japan}

\author[0000-0001-7641-5497]{Itsuki Sakon}
\affiliation{Department of Astronomy, Graduate School of Science, The University of Tokyo, Bunkyo-ku, Tokyo 113-0033, Japan}

\author[0000-0002-8234-6747]{Takashi Onaka}
\affiliation{Department of Physics, Faculty of Science and Engineering, Meisei University, 2-1-1 Hodokubo, Hino, Tokyo 191-8506, Japan}
\affiliation{Department of Astronomy, Graduate School of Science, The University of Tokyo, Bunkyo-ku, Tokyo 113-0033, Japan}

\author[0000-0002-8092-980X]{Peredur M. Williams}
\affiliation{Institute for Astronomy, University of Edinburgh, Royal Observatory, Edinburgh EH9 3HJ, UK}

\author[0000-0002-2287-8151]{Victor I. Shenavrin}
\affiliation{Sternberg Astronomical Institute, Moscow State University, Universitetskij pr., 13, Moscow, 119991, Russia}

\begin{abstract}
We present the detection of a broad 8 $\mu$m feature in newly formed dust around the carbon-rich Wolf-Rayet (WC) binary WR 125 from N-band low-resolution (NL; R$\sim$250) spectroscopy between 7.3--13.6 $\mu$m and N-band (11.7 $\mu$m) and Q-band (18.8 $\mu$m) imaging with Subaru/COMICS in 2019 October. WR 125 is a colliding wind binary (${\rm WC7+O9}$) that exhibited renewed dust formation starting in 2018, $\sim$28 years after its first dust formation episode had been observed. We also compare our infrared photometry with historical observations and revise the dust-formation period of WR 125 to 28.1 years. Archival infrared spectra of five dusty WC stars, WR 48a, WR 98a, WR 104, WR 112 and WR 118, obtained with ISO/SWS are reanalyzed and compared with the WR 125 spectrum to search for a similar feature. We analyze the dusty WC spectra using two different extinction curves to investigate the impact of interstellar extinction correction on the presence and/or properties of the 8 $\mu$m feature. All of the dusty WC spectra dereddened with the two different extinction curves show a broad feature around 8 $\mu$m (FWHM$\sim$1-2 $\mu$m). We suggest that these 8 $\mu$m features seen in the dusty WC spectra are related to the Class C unidentified infrared (UIR) features.  
 
\end{abstract}


\section{Introduction} \label{sec:intro}
Wolf-Rayet (WR) stars are the descendants of massive O-type stars characterized by broad emission line spectra. Typically, they have masses of 10--25 ${\rm \msun}$ and high luminosity of ${\rm L\gtrsim10^{5}\lsun}$ \citep{Crowther07}. They lose mass (${\rm \gtrsim 10^{-5}\msunyr}$) by massive and high-velocity stellar winds (${\rm \gtrsim 1000 km/s}$), which make them lose the hydrogen outer layer, and the material synthesized in the stellar interior and transferred to the surface causes their notable emission spectra \citep{Gamow43}. 

WR stars are classified into WN subtype, which exhibits strong \ce{He} and \ce{N} lines, WC subtype, which exhibits strong \ce{He}, \ce{C}, and \ce{O} lines, and the rare WO subtype, which is similar to WC subtype but shows more dominant \ce{O} lines. Despite their hostile environment, WC stars have been reported to exhibit excess infrared emission that indicates dust formation \citep{Williams87}. 

Dust formation around WR stars is expected to be difficult as discussed by \citet{Hackwell79}; the intense radiation fields of WR stars could prevent dust formation, and carbon is predicted to be ionized and not dense enough to form dust in the homogeneous and spherically symmetric wind of a single star. However, \citet{Williams87} clearly showed that circumstellar dust is formed in the WR wind. \citet{Usov91} investigated the mass-loss process of WR binaries and showed that the collision of the winds from WR 140 and its companion around periastron passage caused strong gas compression, which allows the gas to cool quickly to form dust. The dust-forming region in the wind of massive stars needs to be shielded from the hot stellar photosphere \citep[e.g.][]{Kochanek11} and the shielding is provided in the dense region formed by wind-wind collisions in case of WR binaries.

Dust formation around WC stars can be classified into ``persistent'' or ``episodic/periodic'' formation \citep{Williams19}. Episodic/periodic dust formation occurs in a brief time around periastron passage in long-period WC+OB binary systems, when the pre-shock wind density becomes high \citep{Williams99,Williams09}, while persistent dust makers might be binaries in circular orbits \citep{Williams19}. Infrared imaging of persistent dust makers often reveals continuous ``pinwheel'' structures while that of the episodic dust maker WR 140 shows an expanding arc of dust centered on the star.

WR stars have been previously overlooked as a significant dust source since only the most massive stars were thought to undergo high enough wind-driven mass loss to enter the WR phase. Supernovae (SNe) have therefore been regarded as a dominant source of dust in the early universe \citep{Dwek11, Aleksandra19, Bakx21}. However, interactions with a binary companion can play an important role in the mass loss of massive stars \citep{Marco17} and enable new channels of WR formation even in metal-poor environments. \citet{Sana12} claimed that more than 70\% of all massive stars will exchange mass with a companion, which implies that binary interaction dominates the evolution of massive stars. \citet{Machida08} also theoretically suggested that massive binary systems are more abundant even in the early universe. 

\citet{Lau20a} indeed demonstrated that WC binaries should be important sources of carbon-rich dust at metallicities of $Z\gtrsim 0.5 Z_\odot$. Recently \citet{Lau21} also suggested that early-type WC binaries may have a significant impact on the dust budget in metal-poor galaxies based on the analysis of mid-infrared light curve of extragalactic dust-forming WC binaries obtained with the Infrared Array Camera on Spitzer. WC stars in a binary system, therefore, can potentially become the initial dust producer in the history of the universe, and infrared spectroscopy of dusty WC binaries are important to investigate what kind of dust is formed around them.

Given their C-rich nature, WC binaries could be a plausible source of polycyclic aromatic hydrocarbons (PAHs), which have been widely accepted as a likely candidate of the carriers of the unidentified infrared (UIR) features \citep{Leger84,Allamandola85,Tielens08}. The main stellar source of PAHs has been considered to be asymptotic giant branch (AGB) stars \citep{Galliano08}, while they may be formed from fragmentation of large carbonaceous dust in the interstellar medium \citep[e.g.,][]{Seok14}. However, it takes a longer time for low- to intermediate-mass stars to evolve into the AGB phase compared with the evolution of massive stars to WR stars. Therefore, their injection of PAHs into the ISM is delayed compared with that of massive stars \citep{Galliano08}. The UIR 6.2\,$\mm$ emission feature was directly observed up to ${\rm z=4}$ so far \citep{Riechers14}. Whether the UIR features are present at redshift higher than 6, where there is little or no contribution from low- to intermediate-mass stars is an issue that will be addressed by observations with future infrared space telescopes.

In order to understand the origins of organics as the carriers of the UIR features in the early universe, it needs to be understood whether or not WR stars can produce organics based on the analysis of UIR features detected in a certain evolutionary stage of the WR stars. So far, the presence of the UIR features in some dusty WC stars has been reported; WR 104 and WR 112 exhibit broad features around 8\,$\mm$ \citep{Cohen89}, WR 48a exhibits 6.4 and 7.9\,$\mm$ features \citep{Chiar02}. \citet{Marchenko17} investigated the combined spectra from the Short-Wavelength Spectrometer (SWS) on the Infrared Space Observatory (ISO) and showed that the UIR features are present. Further infrared spectroscopic observations of dusty WC stars are needed to investigate the nature of possible carriers of the UIR features in the early universe.

In this paper, we investigate the re-emerging active dust formation in the colliding-wind WC binary WR 125. WR 125 had been recognized as a WR star by \citet{Iriarte56} and was later classified as a ${\rm WC7+O9}$ system \citep{Williams94}. The distance is estimated to be $3.36_{-0.65}^{+0.99}$ kpc \citep{Rate20}. \citet{Williams92} reported evidence of dust formation in WR 125 based on infrared monitoring observations motivated by the similarities in radio and X-ray properties between WR 125 and WR 140, a famous dusty WR binary. WR125 has also been the target of long-term X-ray monitoring \citep{Midooka19}. 

\citet{Williams19} claimed that WR 125 was found to be an episodic/periodic dust producer whose time interval is $\sim 28.3$ yr. The previous maximum IR flux was in 1992--1993 \citep{Williams94}. The latest IR re-brightening is already proceeding and the periastron passage is thought to have occurred around 2020. Therefore, the year of 2020 is a valuable time to investigate the properties of newly formed dust around WR 125. \citet{Williams94} also reported the result of mid-infrared spectroscopy of WR 125 at periastron in 1992 and pointed out the presence of a feature at 8.6\,$\mm$, which may be related to the UIR features.

This paper is organized as follows. In Section \ref{sec:obs}, we describe the infrared observations of WR 125 and those of five dusty WC stars for comparison with the WR 125 spectrum. Section \ref{sec:result} presents the result of the SED analyses. In Section \ref{sec:discussion}, we discuss the properties of the feature detected in the dusty WC spectra and its possible carriers. Finally, Section \ref{sec:conclusion} summarizes the results and draws conclusions.

\section{Observations} \label{sec:obs}
\subsection{Mid-infrared spectroscopy and imaging of WR 125 with Subaru/COMICS}
The COoled Mid-Infrared Camera and Spectrometer \citep[COMICS;][]{Kataza00,Okamoto03,Sako03} is installed at the Cassegrain focus of the 8.2 meter Subaru Telescope at the top of MaunaKea. COMICS has both imaging and spectroscopic capabilities from 7.5 to 25\,$\mm$ covering the N and Q bands. COMICS enables high sensitivity and excellent spatial resolution mid-infrared observations, owing to the high altitude and large aperture of the Subaru telescope.

The observations of WR 125 with Subaru/COMICS were carried out on 2019 October 11. We carried out N- and Q-band imaging and N-band low-resolution (NL) spectroscopy. The total exposure time of the N- and Q-band imaging is both 200 seconds. The chopping throw was set as 10 arcsec, and both beams were used for the analysis. The flux calibration was made using $\gamma$ Aql (K3II) as a photometric standard star \citep{Cohen99}. The aperture size for the photometry was set to be 2.5 arcsec in radius. IR emission from WR 125 does not appear to be resolved in the N- and Q-band images. The errors of the flux at N and Q are calculated from the images and the error of the standard. 

For the NL spectroscopy, we also set the chopping throw as 10 arcsec. The weather conditions were not ideal and we discarded the data that showed unstable backgrounds. The total effective integration time was 150 sec. The data reduction up to flux calibration was performed by standard procedure for mid-infrared spectroscopic observations with ground-based telescopes \citep{Honda03,Honda04}. The slit throughput depends on the location of the object on the slit and we correct for it. The observed and calibrated spectrum of $\alpha$ Cyg (A2Ia) in the catalogue of spectra obtained with ISO/SWS \citep{Sloan03} were used for the flux calibration for the NL observation of WR 125.

\subsection{Near-infrared imaging of WR 125 with 1.25 m telescope of SAI}
In addition to the data at N (11.7\,$\mm$) and Q bands (18.8\,$\mm$) taken with COMICS, we also used data at J, H, K, L, and M for SED fitting from the 1.25 m telescope of the Sternberg Astronomical Institute's (SAI) Crimean Laboratory of Moscow State University using a JHKLM photometer with an InSb photovoltaic
detector cooled with liquid nitrogen \citep{Shenavrin11}. The data were taken on 2019 October 9, which was almost the same time as the COMICS observations. Several integrations were made on the object with an exposure of 30--60 seconds. The total integration time in each of the JHKL filters was 5--10 minutes and 20--25 minutes in the M filter. The standard star BS7488 was observed twice --- before and after measurements of the object and the accumulation time in each filter was 1--2 minutes. A detailed description of the observing technique and instrumentation parameters is provided by \citet{Shenavrin11}. The stellar magnitudes of a standard star are taken from the catalog by \citet{Johnson66}. The observed magnitudes and fluxes are summarized in Table \ref{tab:WR125_201910}. 

\subsection{Archival infrared photometry from NEOWISE-R}
We used archival W1 (3.4\,$\mm$) and W2 (4.6\,$\mm$) photometry of WR 125 from the Near-Earth Object WISE Reactivation (NEOWISE-R) mission \citep{Mainzer14,NEOWISE_DOI} obtained within twelve days of the COMICS observations and applied the following filters to identify the best quality data: cc\_flags$=$0000, ph\_qual$=$AA, and qual\_frame$=$5. Three photometric measurements of WR 125 passed these filters: two observations were from 2019 September 29 (MJD = 58755.20263 and MJD = 58755.66054) and one was from 2019 October 7 (MJD = 58763.44457). The reported w1mpro and w2mpro magnitudes for the two 2019 September 29 observations were $W1 = 4.782\pm0.060$ and $4.846\pm 0.041$ mag and $W2 = 3.946\pm0.011$ and $3.776\pm0.007$ mag. The reported magnitudes for the 2019 October 7 observations was $W1 = 5.311\pm0.045$ mag and $W2 = 3.565\pm0.006$ mag.

Since the observations were partially saturated, the w1mpro and w2mpro magnitudes of the three observations were individually corrected according to saturation photometric bias corrections from the Explanatory Supplement to the NEOWISE Data Release Products \citep{Cutri15}\footnote{Retrieved 2021 December from \url{https://wise2.ipac.caltech.edu/docs/release/neowise/expsup/sec2_1civa.html}}. When applying the saturation correction, we adopted the largest uncertainty value as the $\pm1\sigma$ uncertainties for the appropriate saturation bias correction values for w1mpro and w2mpro, w1mcorr and w2mcorr. We used an average of these observations, for which a saturation correction was applied individually for our SED analysis of WR 125, where $W1_\mathrm{avg} = 5.68\pm0.12$ mag and $W2_\mathrm{avg} = 4.65\pm0.14$ mag. The magnitudes were converted to fluxes using the zero magnitude flux density and the flux correction for blackbody spectra, $B_{\nu}(800 K)$ \citep{Wright10}. The calculated magnitudes and fluxes are listed in Table \ref{tab:WR125_201910}.

\subsection{Infrared photometry from AKARI and UKIRT in 2006--2008}
We also use two infrared photometric data obtained in 2006--2008 to discuss the photometric period of WR 125 in Sec. \ref{sec:period}. We use archival data from the AKARI Infrared Camera (IRC) mid-infrared all-sky survey from 2006 May to 2007 August obtained with two broad band filters of S9W centered at 9\,$\mm$ and L18W at 18\,$\mm$ \citep{Ishihara10}. We also present unpublished K and L$^{\prime}$ photometry on 2007 July 17 and 2008 July 29 from images taken with the United Kingdom Infrared Telescope (UKIRT) using the 1--5\,$\mm$ UKIRT Imager Spectrometer (UIST) \citep{UIST} in programme U/SERV/1755. The observed fluxes and magnitudes are summarized in Table \ref{tab:WR125_2006}

\begin{deluxetable*}{ccccccccccc}[h]
\tablecaption{The WR 125 magnitude and flux observed in 2019.}
\tablewidth{2.0\linewidth}
\tablehead{ & \multicolumn{5}{c}{1.25 m telescope of SAI} & \multicolumn{2}{c}{NEOWISE-R} & \multicolumn{2}{c}{COMICS} \\
Bands & J & H & K & L & M & W1 & W2 & N & Q} 
\startdata
Dates & \multicolumn{5}{c}{October 9} & \multicolumn{2}{c}{September 29--October 7} & \multicolumn{2}{c}{October 11} \\ 
Wavelength[$\mm$] & 1.25 & 1.63 & 2.19 & 3.47 & 4.7 & 3.37 & 4.62 & 11.7 & 18.8 \\ \hline
Magnitude & 9.28$\pm$0.03 & 8.63$\pm$0.03 & 7.69$\pm$0.03 & 5.58$\pm$0.03 & 4.81$\pm$0.05 & 5.68$\pm$0.12* & 4.65$\pm$0.14* & - & - \\ 
Flux[Jy] & 0.30$\pm$0.01 & 0.37$\pm$0.01 & 0.55$\pm$0.02 & 1.69$\pm$0.05 & 1.94$\pm$0.09 & 1.66$\pm$0.19 & 2.39$\pm$0.31 & 1.55$\pm$0.13 & 0.51$\pm$0.15
\enddata
\tablecomments{*NEOWISE-R magnitudes shown here are the average of three data corrected according to saturation photometric bias corrections, which were obtained on 2019 September 29 and October 7. \label{tab:WR125_201910}}
\end{deluxetable*}

\begin{deluxetable*}{ccccccc}[h]
\tablecaption{The WR 125 magnitude and flux observed in 2006--2008.}
\tablewidth{2.0\linewidth}
\tablehead{& \multicolumn{2}{c}{AKARI} & \multicolumn{4}{c}{UKIRT} \\
Bands & S9W & L18W & K & L$^{\prime}$ & K & L$^{\prime}$} 
\startdata
Dates & \multicolumn{2}{c}{2006 May--2007 August} & \multicolumn{2}{c}{2007 July 17} & \multicolumn{2}{c}{2008 July 29} \\ 
Wavelength[$\mm$] & 9 & 18 & 2.2 & 3.8 & 2.2 & 3.8 \\ \hline
Magnitude & - & - & 8.20$\pm$0.04 & 7.81$\pm$0.03 & 8.16$\pm$0.05 & 7.76$\pm$0.03 \\ 
Flux[Jy] & 0.59$\pm$0.07 & 0.54$\pm$0.12 & - & - & - & -
\enddata
\tablecomments{The fluxes in S9W and L18W filters are the average of six and four scans, respectively. \label{tab:WR125_2006}}
\end{deluxetable*}
 
\subsection{Archival ISO/SWS spectra of dusty WR stars}
The 2.4--45\,$\mm$ medium resolution ($\lambda/\Delta\lambda\approx$250--600) infrared spectra of five dusty WC stars, WR 48a, WR 98a, WR 104, WR 112, and WR 118, obtained with ISO/SWS \citep{Hucht96} were reanalyzed and compared with the WR 125 NL spectrum. In this paper, we use the 2.4--20\,$\mm$ region of the spectra for our analysis. The observations were made between February and March 1996. Details of the observations are provided by \citet{Hucht96}. The stitched spectra of the five dusty WC stars were downloaded from the ISO/SWS spectral database processed by \citet{Sloan03} and used for the analysis. The ISO/SWS spectra of these five WC stars were examined in previous studies \citep[e.g.][]{Chiar02, Marchenko17}. In this study, we examine the presence of emission features using the same analytical technique that we apply to the WR 125 NL spectrum as described in Sec. \ref{sec:result}.

\section{Results} \label{sec:result}
\subsection{The infrared photometric period of WR 125} \label{sec:period}
\begin{figure*}[htbp]
\centering
\includegraphics[width=\linewidth]{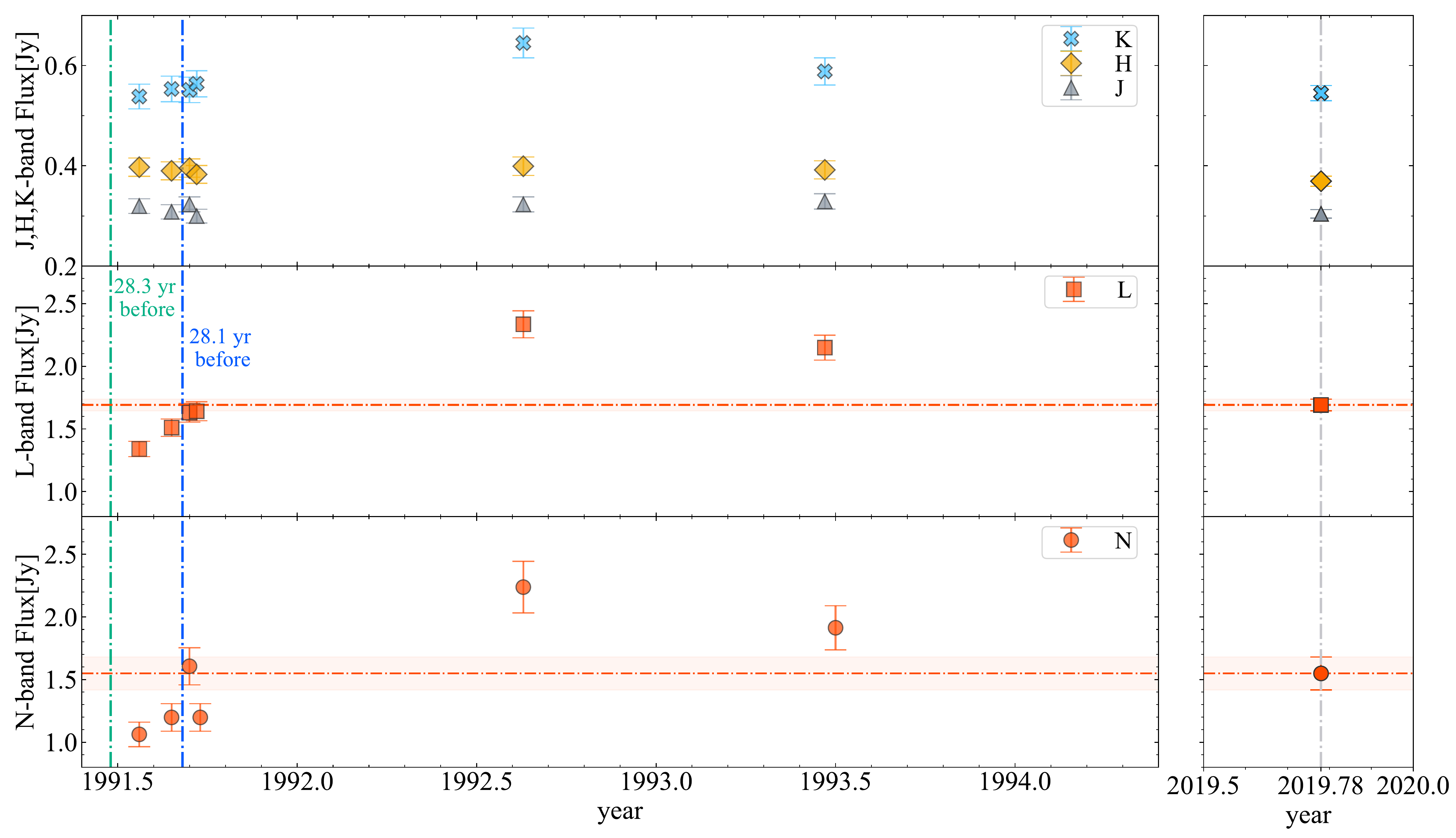}
\caption{Comparison of J, H, K, L,and N-band flux in our data taken in 2019.78 with those in 1991--1994 \citep{Williams92, Williams94}. The green and blue lines show 1991.48 and 1991.68, which are 28.3 and 28.1 yr before our observations, respectively. The data in 1991.56, 1991.65, and 1991.72--73 were taken by the ESO 3.6m telescope with the filters centered at 1.24, 1.64, 2.18, 3.76 and 12.9\,$\mm$, while those in 1991.70, 1992.63, and 1993.47--50 were taken by the UKIRT with the filters centered at 1.25, 1.64, 2.20, 3.75, and 12.5\,$\mm$. We note that these wavelengths are slightly different from our observations at 1.25, 1.63, 2.19, 3.47 and 11.7\,$\mm$. \label{fig:WR125_period}}
\end{figure*}

WR 125 is an episodic/periodic dust maker and the interval between the episodes was suggested to be $\sim$28.3 yr based on the comparison of 2018 NEOWISE-R observations with the L$^{\prime}$ and M photometry in 1990--1991 \citep{Williams19}. We compared the ground-based J, H, K, L, and N photometry from 2019 October with those in 1990--1993 to estimate the infrared period of WR125. Figure \ref{fig:WR125_period} shows the near-mid infrared flux in 1991--1993 \citep{Williams92, Williams94} and that of our observations. The J- and H-band fluxes do not show any appreciable variation between 1990 to 2019. The L-and N-band data taken in the 2019.78 observations are 26 and 46 \% brighter, respectively, than those taken at 1991.56, which is slightly discrepant with the suggested 28.3 yr periodicity of WR 125 mid-IR light curve \citep{Williams19}. However the 2019.78 observations agree with those at 1991.70, suggesting that the interval of the dust formation episodes may be $\sim$28.1yr rather than 28.3 yr.

In their discussion of the period, \citet{Arora21} did not rule out the possibility of $\sim$14 yr, i.e. half this length. Infrared photometric data obtained with AKARI/IRC and UKIRT/UIST in 2006--2008 can be used to rule out a $\sim14$ yr period. The WR 125 fluxes in the AKARI S9W and L18W filters are 0.59$\pm$0.07 and 0.54$\pm$0.12 Jy, which are the average of six and four scans, respectively. Observations taken with UKIRT at 8.7 and 19.5\,$\mm$ in 1992.63, $\sim$14 yr before 2006--2007, measure fluxes of 3.2 and 0.9 Jy, respectively \citep{Williams94}. This difference makes it highly unlikely that WR 125 exhibits dust formation with a $\sim$14 yr period.

The K magnitudes of WR 125 observed with UKIRT are 8.20$\pm$0.04 and 8.16$\pm$0.05 mag in 2007.54 and 2008.58, respectively, while L$^{\prime}$ magnitudes are 7.81$\pm$0.03 and 7.76$\pm$0.03 mag. These magnitudes are also significantly fainter than K and L$^{\prime}$ magnitudes, 7.60 and 5.16 mag, respectively, taken at 1993.47 at UKIRT \citep{Williams94}\footnote{\citet{Williams94} state that the uncertainties of the K and L$^{\prime}$ magnitudes are 0.05 mag or smaller}. The AKARI and UKIRT observations in 2006--2008 and the IR photometry from \citet{Williams94} rule out the possibility of a $\sim$14 yr period.

\subsection{The spectral energy distribution of WR 125} \label{sec:SED}
\subsubsection{Extinction and free-free emission correction}
We fitted SEDs to both photometric and NL spectrophotometric data of WR 125 obtained in 2019 (i.e., the photometric data in Table \ref{tab:WR125_201910}) to model the continuum emission. The first step is to correct for interstellar extinction. \citet{Weingartner01} provided an extinction curve based on their dust model, while \citet{Lutz96} suggested excess extinction at 5--8\,$\mm$ based on the HI recombination lines towards the Galactic center. The results of \citet{Lutz96} were confirmed by later observations by \citet{Indebetouw05} and \citet{Nishiyama09}. \citet{Chiar06} combined extinction points from \citet{Indebetouw05} and \citet{Lutz99} to represent the continuum extinction and derived an extinction curve for the local ISM along with the silicate profiles of WR 98a, assuming that the WR 98a spectra exhibit featureless continuum. Most recently, \citet{Gordon21} thoroughly analyzed the extinction based on Spitzer observations and derived the extinction curve for the diffuse ISM, which is in between \citet{Weingartner01} and \citet{Indebetouw05} in terms of the excess in the MIR.

Taking account of the uncertainties in the MIR extinction curve and the large extinction for the WR stars in the present study, we use two extinction curves to estimate their effects on the resultant spectra. One is the synthetic extinction curve for $R_{V}=3.1$, which is an average value of diffuse gas in Milky Way, developed by \citet{Weingartner01} (hereafter WD01). The WD01 extinction curve is most widely employed in past analyses and thus makes comparison easy. The other is the average extinction curve recently derived by \citet{Gordon21} (hereafter G21). We utilize both the WD01 and G21 extinction curves because they were derived from independent techniques and therefore strengthen the robustness of the SED analysis after extinction correction.

Both the WD01 and G21 extinction curves are for diffuse sight lines. The ice absorption feature at 3.1\,$\mm$ is not seen in all the ISO/SWS WC spectra, while the WR\,112 spectrum shows an absorption band of \ce{CO2} ice at 4.27\,$\mm$, suggesting that the sight line may pass through a dense cloud. The 9.7\,$\mm$ silicate feature strength for dense sight lines is smaller than that for diffuse sight lines per $A_{V}$, but the shape does not change appreciably \citep{Gordon21}. The absorption in WR\,112 spectrum can be overcorrected by using the extinction curves for diffuse sight lines. However, it will not heavily affect the overall shape of the spectra after extinction correction.

\begin{figure*}[htbp]
\centering
\includegraphics[width=\linewidth]{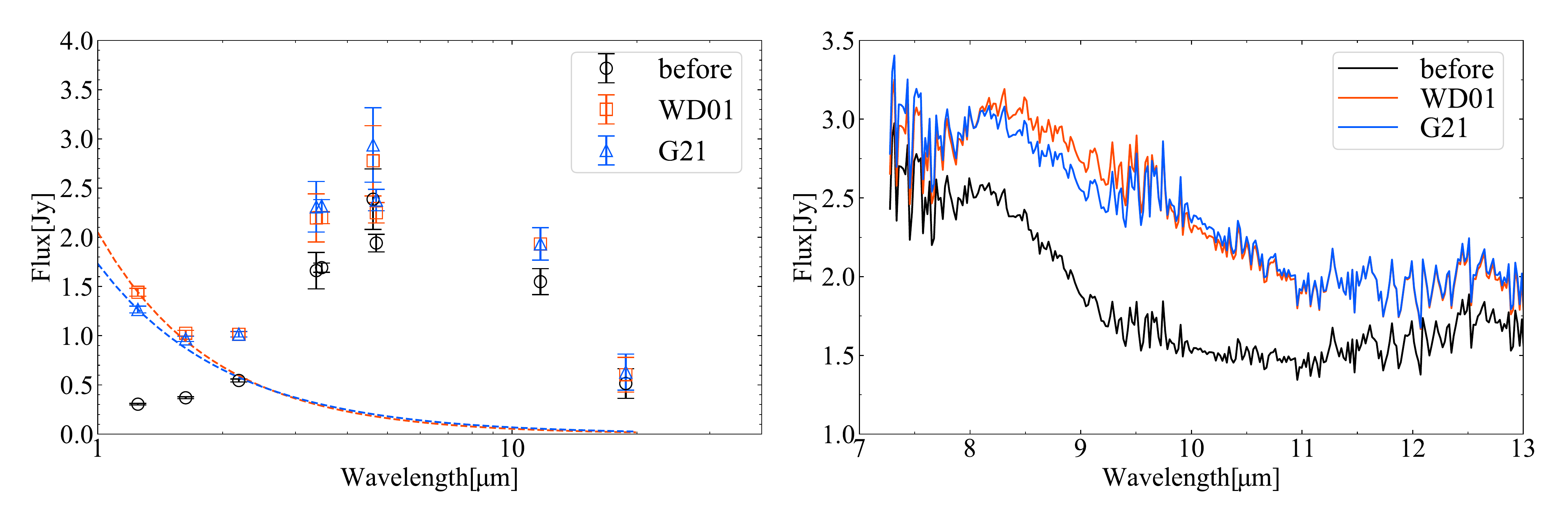}
\caption{(left)J, H, K, L, W1, W2, N and Q-band flux obtained in October before and after extinction correction using WD01 and G21 extinction curves with $A_{V}=5.89$. The flux after extinction correction shows excess in near-infrared region. The dashed lines show the power-law for free-free emission correction derived using the photometric data in \citet{Williams92}. The power-law is normalized by setting the J-band flux after the correction as zero for each $A_{V}$ value. The x-axis is set in a log scale to make it easy to distinguish the values. (right) NL spectra of WR125 before and after extinction correction.  \label{fig:WR125_ext}}
\end{figure*}

The $A_{V}$ value of WR 125 is calculated as $5.89\pm0.75$ from $A_{v}=6.48\pm0.83$ \citep{Rate20} using the relation $A_{v}=1.1A_{V}$ \citep{Turner82}\footnote{$V$ indicates the standard Johnson V-band filter, while $v$ indicate the Smith $v$ filter \citep[][; $\lambda_{C}=5160$\AA\ chosen to avoid the strongest WR emission lines.]{Smith68}}. Figure \ref{fig:WR125_ext} shows the photometric data and NL spectra before and after extinction correction with $A_{V}=5.89$ using WD01 and G21 extinction curves.

The flux after extinction correction shows excess in the near-infrared region, which is attributed to the free-free emission. To extract the contribution from dust in the near-infrared, we estimate the free-free emission component using the photometric data of \citet{Williams92} taken in 1981--1989, before the dust-formation episode beginning in 1990, when WR 125 seems to not show a sign of an appreciable amount of dust formed. We approximate the contribution from the free-free emission as a power-law function and assume that the J-band flux at present does not have any dust emission as shown in Figure \ref{fig:WR125_ext} (left). Table \ref{tab:corrected_flux} shows the flux at each band after extinction and free-free emission correction with different extinction corrections. A free-free emission correction is also applied for each NL spectrum after extinction correction and, it does not affect the shape of the spectra.

\begin{deluxetable*}{cccccccccc}[h]
\tablecaption{The WR 125 flux after extinction and free-free emission correction.}
\tablewidth{2.0\linewidth}
\tablehead{ & \multicolumn{4}{c}{1.25 m telescope of SAI} & \multicolumn{2}{c}{NEOWISE-R} & \multicolumn{2}{c}{COMICS} \\
& H & K & L & M & W1 & W2 & N & Q \\
Wavelength[$\mm$] & 1.63 & 2.19 & 3.47 & 4.7 & 3.37 & 4.62 & 11.7 & 18.8} 
\startdata
Flux(WD)[Jy] & 0.08$\pm$0.03 & 0.42$\pm$0.03 & 1.91$\pm$0.06 & 2.07$\pm$0.10 & 1.90$\pm$0.24 & 2.59$\pm$0.36 & 1.89$\pm$0.17 & 0.58$\pm$0.18 \\ 
Flux(G21)[Jy] & 0.10$\pm$0.03 & 0.44$\pm$0.03 & 2.02$\pm$0.06 & 2.18$\pm$0.11 & 2.00$\pm$0.26 & 2.74$\pm$0.38 & 1.88$\pm$0.17 & 0.60$\pm$0.18
\enddata
\tablecomments{The contribution from the free-free emission is approximated as a power-law function and the J-band flux at the present observations is assumed not to have any dust emission as shown in Figure \ref{fig:WR125_ext} (left). \label{tab:corrected_flux}}
\end{deluxetable*}

\subsubsection{Dust continuum fitting}
\begin{figure*}[htbp]
\centering
\includegraphics[width=\linewidth]{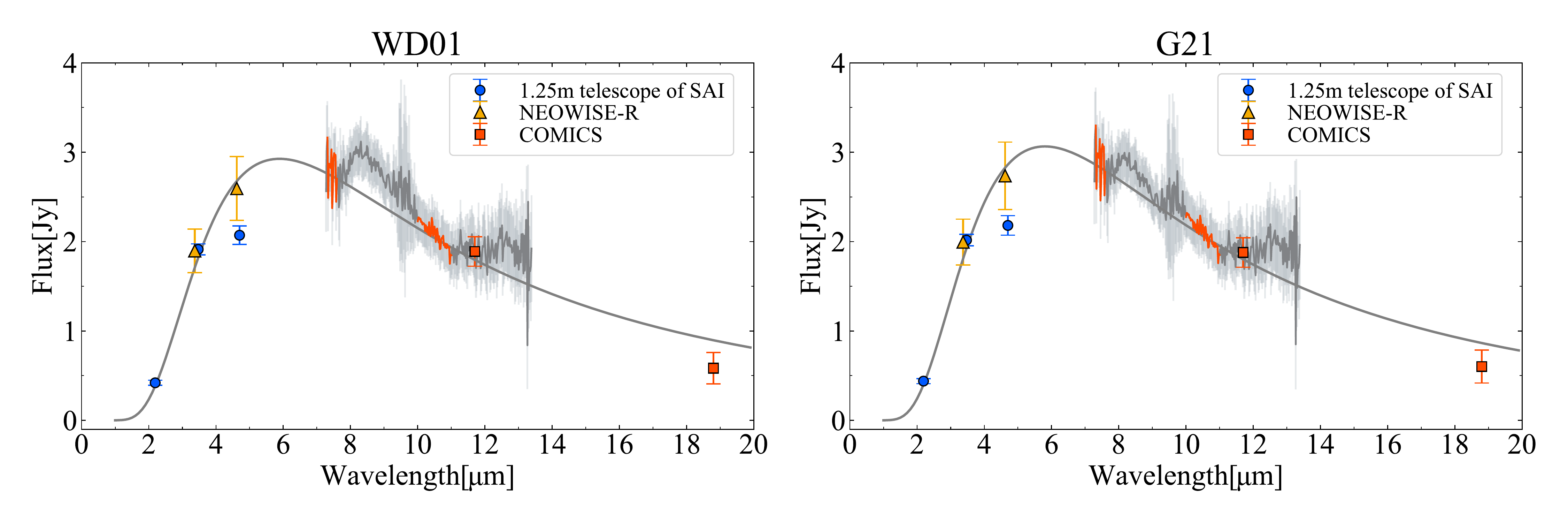}
\caption{One-temperature modified black body (MBB) fitting of WR125 data after extinction and free-free emission correction. The 7.3--7.6 and 10--11\,$\mm$ regions in NL data are used for the fitting in addition to K, L, W1, W2, N, and Q-bands photometric data. The grey lines show the uncertainties of the WR 125 NL spectra.
\label{fig:WR125_fit}}
\end{figure*}

The continuum emission from dust is approximated by a modified blackbody (MBB):
\begin{equation}
    F_{\nu} \propto \nu^{\beta}B_{\nu}(T) \propto \lambda^{-\beta}B_{\nu}(T),
\end{equation}
where $\nu^{\beta}$ is the dust emissivity approximated by a power-law and $B_{\nu}(T)$ is the Planck function for the grain temperature $T$. We exclude the J- and H-bands from the fitting, assuming that dust emission is negligible at these wavelengths. We also incorporate NL data points between 7.3--7.6 and 10.0--11.0\,$\mm$, which are selected to avoid strong terrestrial atmospheric attenuation and not include known features, to estimate the shape of the continuum between the M and N bands. We use an isothermal model on the assumption that the dust would not have had time to expand far from the stars and take up a significant range of radiative equilibrium temperatures. We searched for the best fitting value for $\beta$ from 0.0 to 2.0 with a step of 0.1 and chose $\beta=0.2$ and $0.3$ for the spectra dereddened with the WD01 and G21 extinction curves, respectively. The values are obviously smaller than those obtained in laboratories \citep[$\beta=$1--2;][]{Rouleau91} but are not totally unreasonable given that the circumstellar dust around WR 125 likely exhibits more complicated properties than can be characterized by a single temperature. The goal of this analysis is to characterize the shape of the continuum as opposed to accurately measuring the dust properties, which can be provided by a simple MBB approximation. The best fitting parameters are listed in Table \ref{tab:WR_gaussfit}. Figure \ref{fig:WR125_fit} shows the result of MBB fitting with the two different extinction curves. The ground-based M-band photometric point deviates from the fitted MBB. However, the overall shape of the fit is consistent with all other photometric points including the NEOWISE-R W2 measurement. The spectra of WR 125 after the two extinction corrections both show a broad feature around 8\,$\mm$.

\subsection{ISO/SWS spectra of other dusty WC stars}
In order to investigate whether or not the 8\,$\mm$ feature seen in WR 125 is present in other dusty WC stars, we reanalyzed the ISO/SWS spectra of five dusty WC stars: WR 48a, WR 98a, WR 104, WR 112 and WR 118. Extinction corrections were applied to these five dusty WC stars in the same way as for WR 125 using the WD01 and G21 extinction curves. The $A_{V}$ values and their uncertainties for WR 48a, WR 104, and WR 118 were calculated from $A_{v}$ measured by \citet{Rate20}. We adopted $A_{v}$ values from \citet{Hucht01} for WR 98a and WR 112 because of the large distance and extinction uncertainties estimated for these sources by \citet{Rate20}. Additionally, applying the $A_{v}$ values from \citet{Rate20} for WR 98a and WR 112 does not adequately correct for the interstellar 9.7\,$\mm$ silicate absorption feature. The $A_{V}$ of WR 98a is estimated as 12.54 from $A_{v}=13.79$ from \citet{Hucht01}, and this is the only value listed in the literature to our knowledge. The mean $A_{V}$ value of WR 112 is estimated as 11.13 from $A_{v}=12.24$ from \citet{Hucht01}. Other $A_{v}$ values are reported for WR 112 in \citet{Hucht01} and we set the difference between maximum/minimum value in the table and the mean value as the uncertainties of the $A_{V}$ of WR 112. Table \ref{tab:WR_gaussfit} lists the $A_{V}$ values and uncertainties of the WC stars. The spectra shown in the figures are reduced using the mean $A_{V}$ values. 

\begin{figure*}[htbp]
\centering
\includegraphics[width=\linewidth]{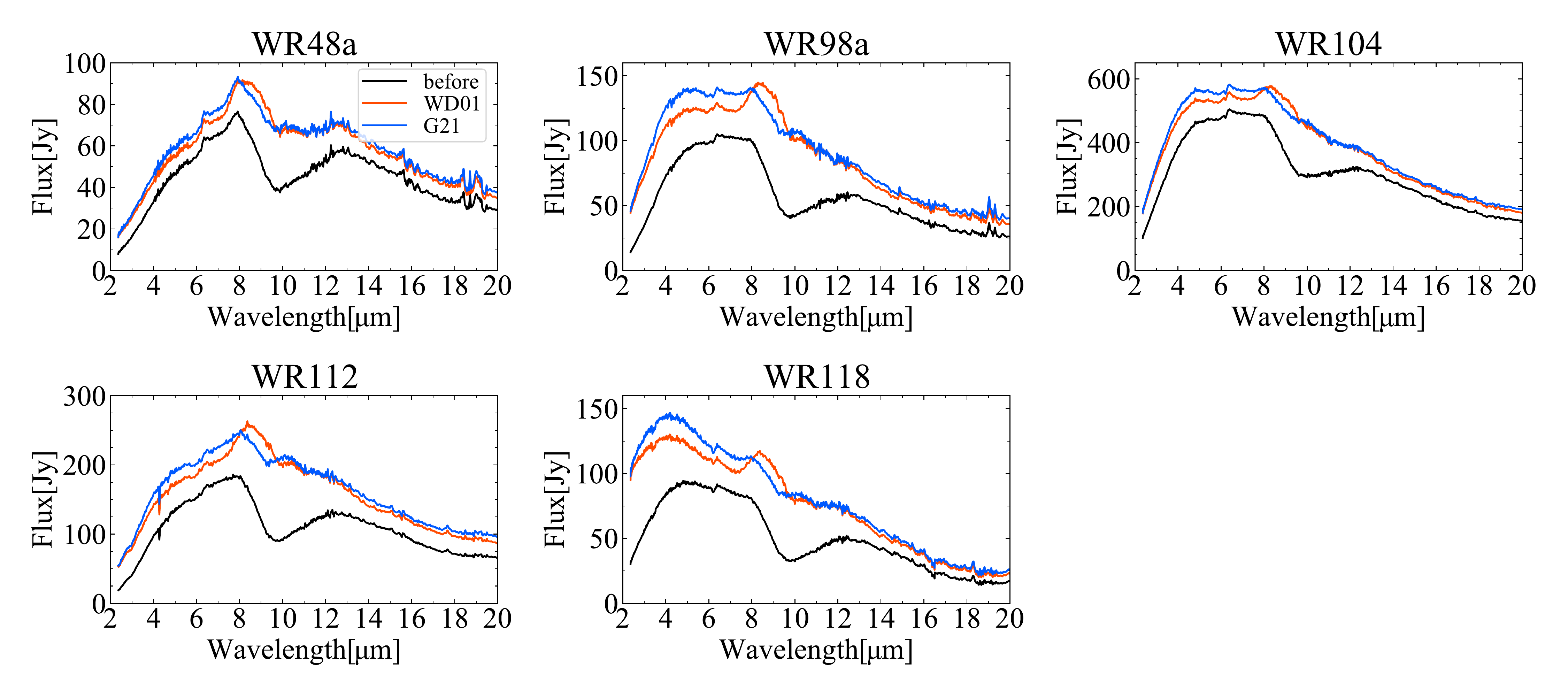}
\caption{ISO/SWS spectra of WR 48a, WR 98a, WR 104, and WR 112 before and after extinction correction using WD01 and G21 extinction curves. The original spectra before the correction are downloaded from the ISO/SWS spectral database processed by \citet{Sloan03}. $A_{V}$ values used for the correction are 7.53, 12.54, 6.06, 11.13, and 12.44, respectively. \label{fig:ISO_ext}}
\end{figure*}

Figure \ref{fig:ISO_ext} shows the ISO/SWS spectra of the five WC stars before and after extinction correction using the WD01 and G21 extinction curves. The extinction correction from either curve adequately corrects the deep 9.7\,$\mm$ silicate absorption feature in each spectrum. All of the ISO/SWS WC spectra after correction show possible emission features at around 6.4 and 8\,$\mm$.

To model the continuum emission, a two-temperature MBB is used for ISO/SWS WC spectra. The model can be characterized by the following equation:
\begin{equation}
a_{1}\lambda^{-\beta}B_{\nu}(T_{1})+a_{2}\lambda^{-\beta}B_{\nu}(T_{2}).   
\end{equation}

\begin{figure*}[htbp]
\centering
\includegraphics[width=\linewidth]{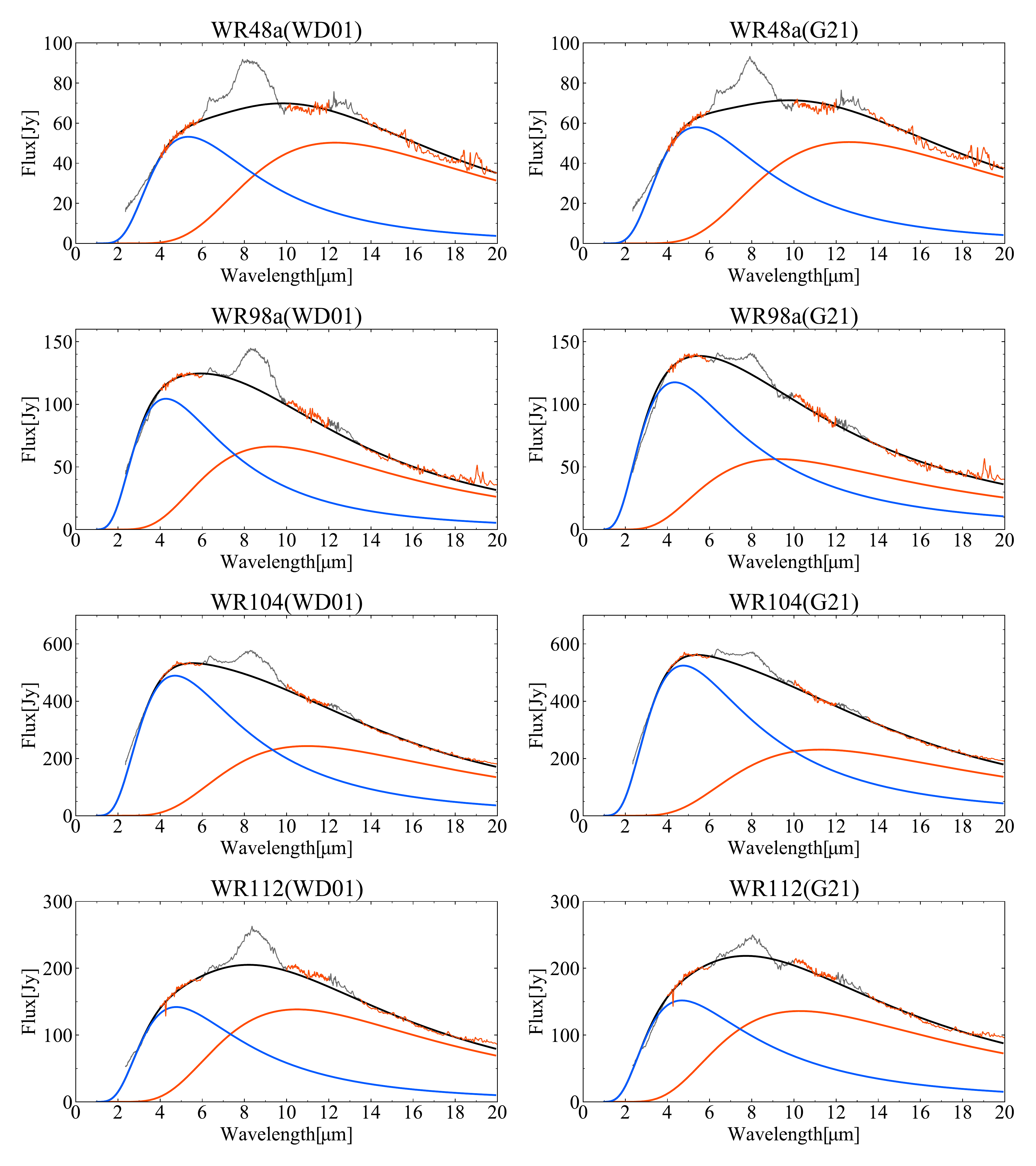}
\caption{Two-temperature modified black body (MBB) fitting of WR 48a, WR 98a, WR 104, and WR 112 data after extinction correction. The 4.0--6.0, 10.0--12.0, and 13.5--20.0\,$\mm$ regions are used for the fitting avoiding the features and shortest wavelength with larger free-free emission contribution. The spectrum of WR 118 is not well fitted by two-temperature MBB fitting with $\beta=0.0-2.0$. \label{fig:ISO_fit}}
\end{figure*}

We chose the continuum points by avoiding the features and near-infrared regions with larger contribution from stellar continuum and/or free-free emission. The $\beta$ values were searched for from 0.0 to 2.0 with a step of 0.1 and the best-fitting values for WR 48a, WR 98a, WR 104 and WR 112 are listed in Table \ref{tab:WR_gaussfit}. The spectrum of WR 118 cannot be fitted well by the two-temperature MBB model with $\beta=0.0-2.0$. Despite the spectrum exhibiting possible 6.4 and 8\,$\mm$ features, we exclude WR 118 from the following analysis because it is difficult to fit a satisfactory continuum model. Figure \ref{fig:ISO_fit} shows the result of the continuum fitting to the extinction corrected spectra of WR 48a, WR 98a, WR 104 and WR 112. 

\begin{deluxetable*}{ccccccccccccc}[h]
\tablecaption{Properties of the dusty WC stars and best-fit parameters.}
\tablewidth{2.0\linewidth}
\tablehead{ & & & \multicolumn{5}{c}{WD01} & \multicolumn{5}{c}{G21} \\ 
WR & Spectral Type & $A_{V}$ & $a_{1}/a_{2}$ & $T$[K] & $\beta$ & peak[$\mm$] & FWHM[$\mm$] & $a_{1}/a_{2}$ & $T$[K] & $\beta$ & peak[$\mm$] & FWHM[$\mm$]}
\startdata
125 & WC7ed+O9III$^{1,2}$ & 5.89$\pm$0.75$^{9}$ & - & 802 & 0.2 & $8.49^{+0.24}_{-0.25}$ & $1.63^{+0.74}_{-0.60}$ & - & 786 & 0.3 & $8.27^{+0.33}_{-0.30}$ & $1.16^{+0.89}_{-0.87}$  \\ 
48a & WC8vd+WN8$^{3,2}$ & 7.53$\pm$0.64$^{9}$ & 62 & 235 & 2.0 & $8.18^{+0.06}_{-0.05}$ & $1.86^{+0.07}_{-0.07}$ & 63 & 229 & 2.0 & $7.91^{+0.01}_{-0.01}$ & $1.60^{+0.03}_{-0.03}$ \\ 
& & & & 542 & & & & & 539 & & &  \\ 
98a & WC8-9vd+?$^{4,2}$ & 12.54$^{10}$ & 22 & 346 & 1.5 & 8.58 & 1.54 & 9 & 408 & 0.9 & 8.08 & 1.14 \\ 
& & & & 757 & & & & & 867 & & &  \\ 
104 & WC9d+B0.5V$^{5}$ & 6.06$\pm$0.62$^{9}$ & 21 & 303 & 1.4 & $8.50^{+0.12}_{-0.11}$ & $1.82^{+0.05}_{-0.08}$ & 18 & 301 & 1.3 & $8.08^{+0.04}_{-0.03}$ & $1.38^{+0.03}_{-0.03}$ \\ 
& & & & 706 & & & & & 715 & & &  \\ 
112 & WC8-9d+OB?$^{6,7}$ & $11.13^{+0.87}_{-4.81}$ $^{10}$ & 36 & 312 & 1.5 & $8.52^{+0.11}_{-0.39}$ & $1.58^{+0.05}_{-0.36}$ & 22 & 360 & 1.0 & $8.04^{+0.03}_{-0.08}$ & $1.08^{+0.02}_{-0.03}$ \\ 
& & & & 689 & & & & & 792 & & &  \\ 
118 & WC9d$^{8}$ & 12.44$\pm$0.65$^{9}$ & - & - & - & - & - & - & - & - & - & -
\enddata
\tablecomments{The references are 1---\citet{Williams94}; 2---\citet{Williams19}; 3---\citet{Zhekov14}; 4---\citet{Cohen91}, 5---\citet{Williams00}; 6---\citet{Massey83}; 7---\citet{Lau20b}; 8---\citet{Conti90}; 9---\citet{Rate20}; 10---\citet{Hucht01}. The WR 125 spectrum is fitted by an isothermal MMB model characterized by $a\lambda^{-\beta} B_{\nu}(T)$, while the ISO/SWS WC spectra are fitted by a two-temperature MMB model characterized by $a_{1}\lambda^{-\beta} B_{\nu}(T_{1})+a_{2}\lambda^{-\beta} B_{\nu}(T_{2})$, where $T{1}<T_{2}$. The $a_{1}/a_{2}$ and $T$ values shown here are those obtained for the extinction corrected spectra using the central $A_{V}$ values. The uncertainties of the peak positions and FWHMs of WR 98a spectra are not shown in the table because they are less than 0.01\,$\mm$. 
\label{tab:WR_gaussfit}}
\end{deluxetable*}

\begin{figure*}[htbp]
\centering
\includegraphics[width=\linewidth]{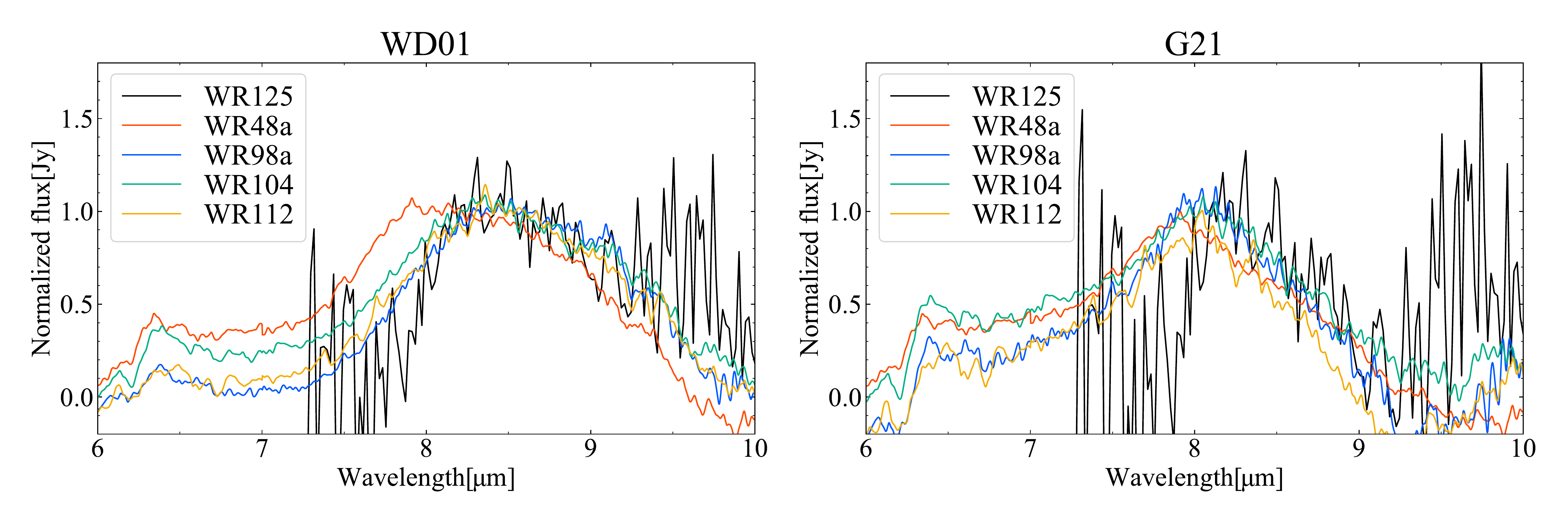}
\caption{Comparison of the WR 125 NL spectrum with four ISO/SWS dusty WR spectra after continuum subtraction. All of the WC spectra show a broad feature in 8\,$\mm$ and the ISO/SWS WR spectra also show a feature in 6.4\,$\mm$. The spectra are normalized around the peak of the 8\,$\mm$ features. The larger noise in 9--10$\mm$ region of the WR 125 spectrum is due to atmospheric absorption. The normalization process makes the noise in the right panel larger than that in the left panel. \label{fig:WR_comparison}}
\end{figure*}

\subsection{Comparison of WR 125 spectra with other dusty WC stars}\label{sec:comparison}
Figure \ref{fig:WR_comparison} compares the WR 125 NL spectra with those of the ISO/SWS WC spectra after continuum subtraction, where all of them are normalized around the peak of the 8\,$\mm$ features. Although the shapes of the spectra vary with the applied extinction curve, all of them exhibit a broad feature around 8\,$\mm$. 

To characterize the 8\,$\mm$ feature in each spectrum, we fit a Gaussian for the feature. The peak positions and FWHMs of the 8\,$\mm$ features for the best fits are shown in Table \ref{tab:WR_gaussfit}, where the error bars account for the uncertainties in $A_{V}$. The 8\,$\mm$ feature in the spectra analyzed with the WD01 extinction curve shows longer peak wavelengths and broader features compared to those dereddened with the G21 curve. We found that larger $A_{V}$ values for the extinction correction result in longer peak wavelengths and vice versa for all of the spectra analyzed using both WD01 and G21 extinction curves. The same trend was observed for FWHMs, where the feature in the spectra analyzed with larger $A_{V}$ results in wider FWHMs.

The shapes of the 8\,$\mm$ feature in the four ISO/SWS WC spectra are very similar to each other, though that of WR 48a appears at a somewhat shorter wavelength. The WR 125 feature deviates from those in the other WC spectra at wavelengths shorter than 8\,$\mm$, which could be attributed to the large uncertainty of the 7--8\,$\mm$ region from ground-based COMICS NL spectra (see Figure \ref{fig:WR125_fit}). Additionally, the lack of the photometric data between 5 and 8\,$\mm$, where the dust emission peaks, contributes to the uncertainties in the dust continuum fit and subtraction. Despite these discrepancies, the slope of the WR 125 feature at wavelengths longer than 8\,$\mm$ agrees well with those of the other WC stars, whichever extinction curve is used. We, therefore, conclude that the 8\,$\mm$ features in the COMICS spectrum of WR 125 and the ISO/SWS spectra of the four WC stars have a similar origin. 

\section{Discussion} \label{sec:discussion}
\subsection{Classification of the 8 $\mu$m feature observed in the dusty WC stars}
Previous studies have already shown the presence of a broad emission feature around 8\,$\mm$ in spectra of the dusty WC stars (WR 104 and WR 112; \citealt{Cohen89}, WR 48a; \citealt{Chiar02}, and a combined spectrum of WR 48a, WR 98a, WR 104, WR 112, and WR 118; \citealt{Marchenko17}), which may be related to the UIR features. The 6.4\,$\mm$ feature is also seen in the ISO/SWS spectra in Figure \ref{fig:WR_comparison} and supports the claim that the WC stars exhibit UIR emission.

With respect to WR 125, \citet{Williams94} indicated the presence of a weak feature around 8.6\,$\mm$ in spectra taken with UKIRT in 1992 October and suggested that the feature may be due to the PAH feature at that wavelength. The peak position of the feature is consistent with the UIR 8.6\,$\mm$ feature, and its width is narrower compared with the broad 8\,$\mm$ feature in the present spectrum of WR 125. A mid-infrared spectrum of WR 125 was also presented by \citet{Smith01}, taken with the Palomar Hale 5 m telescope in 1998 July, and although it is not explicitly mentioned in their work, there is evidence of a feature around 8.4\,$\mm$. The peak position is similar to that in the present spectrum of WR 125, while the band width is still narrower. 

Since WR 125 is an episodic/periodic dust producer, it could be possible that the profile of the 8\,$\mm$ feature varies over time after the dust-formation episode around periastron passage. For example, the evolution of such mid-IR features has been observed in spectra of dusty classical novae after their outburst by \citet{Helton11}, which could also provide new information on the UIR emission in WR stars. However, the peak position and the band width in the past observations can be affected by the extinction correction as discussed in section \ref{sec:comparison}. \citet{Williams94} used the extinction curve in \citet{Rieke85}, which is different from the extinction curves in the present paper, while \citet{Smith01} did not apply extinction correction. Therefore, a possible variation in the properties of the 8\,$\mm$ feature among different orbital epochs is not discussed in the present paper.

The UIR spectra observed in the dusty WC stars can be studied according to the classification defined by \citet{Peeters02}. They classified the observed spectra into Class A, Class B and Class C depending on the peak positions of 6.2 and 7.7\,$\mm$ features. The Class A and Class B UIR spectra show the 8.6\,$\mm$ feature in addition to the 7.7\,$\mm$ feature, while Class C spectra show a broad feature around 8\,$\mm$. The Class D spectra defined by \citet{Matsuura14} are characterized by a broader feature peaking at 6.24\,$\mm$, and a broad single feature from 7 to 9\,$\mm$ peaking at about 7.7\,$\mm$. The UIR spectra from the dusty WC stars do not appear consistent with Class A, which is the dominant class of the UIR spectra seen in the photodissociation regions (PDR) around hot and young stars \citep{Peeters02} as well as in the general interstellar medium \citep{Onaka96,Mattila96}.

\begin{figure*}[htbp]
\centering
\includegraphics[width=\linewidth]{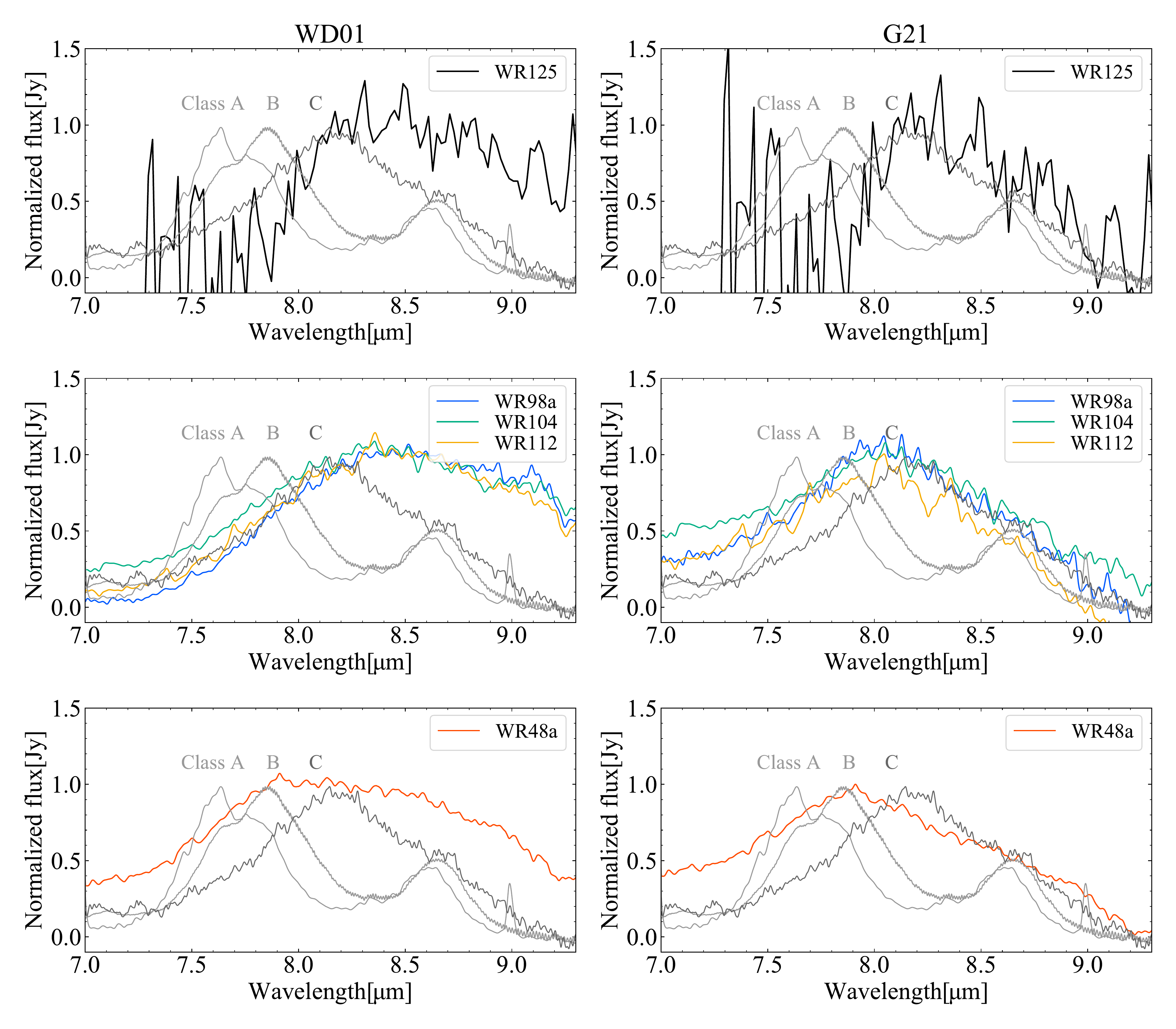}
\caption{(top)Comparison of representative Class A, B, and C UIR spectra with the 8\,$\mm$ feature from the WR 125 NL spectrum. All of them are normalized around the peak. (middle) Comparison with the 8\,$\mm$ features from the ISO/SWS spectra of WR 98a, WR 104, and WR 112. (bottom) Comparison with the 8\,$\mm$ feature from the ISO/SWS spectrum of WR 48a. \label{fig:WR_UIR}}
\end{figure*}

\citet{Chiar02} compared the reduced ISO/SWS WR 48a spectra with the representative spectra of each UIR spectral class shown by \citet{Peeters02} and attributed the 6.4\,$\mm$ and 7.9\,$\mm$ features to Class C and Class B, respectively, based on their peak positions. The broad 8\,$\mm$ features in the WC spectra analyzed in this paper are also compared with those in the same representative spectra of each UIR spectral class in Figure \ref{fig:WR_UIR}. All of the spectra are normalized around the peaks of the features. 

The representative peak wavelengths of the UIR 7.7\,$\mm$ complex are 7.6, 7.8, and 8.22\,$\mm$ for the Class A, Class B and Class C, respectively, as shown by \citet{Peeters02}. The 8\,$\mm$ features of the WC spectra dereddened with the WD01 curve show a longer-wavelength peak than the 8.22\,$\mm$ Class C feature except for that of WR 48a, which peaks at around 8.2\,$\mm$. The widths of the 8\,$\mm$ feature in the WC spectra with the WD01 curve look broader than that of the Class C spectrum. 

While the slope at shorter wavelengths than 8\,$\mm$ is similar to that of the Class C, the slope at longer wavelengths is less steep, which is likely due to the correction of the silicate absorption profile with the WD01 curve. In contrast to the spectra derredened with the WD01 curve, the slopes of the spectra derredened with the G21 curve longer than 8\,$\mm$ are similar to that of the Class C spectrum, while the spectra show excesses at shorter wavelengths, except for WR 125. We cannot exclude the possibility that these excesses are due to the uncertainties in the extinction correction.  

The peak position of the WR 125 spectrum dereddened with the G21 curve is close to that of the Class C spectrum, while those of WR 98a, WR 104, and WR 112 occur at shorter wavelengths than that of the Class C feature. The 8\,$\mm$ feature from WR48a is close to the Class B feature, which is consistent with the classification by \citet{Chiar02}. The peak positions of the ISO/SWS WC spectra are between Class B and Class C \citep[Class B/C;][]{Keller08}. However, the 8.6\,$\mm$ UIR feature is notably absent from the WC spectra (see Figure \ref{fig:WR_UIR}).

Further careful study of the mid-infrared interstellar extinction correction is needed to investigate the accurate shape of the feature. Of the two extinction curves used in our analysis, the G21 extinction curve may lead to more accurate spectral shapes because it is constrained specifically in the mid-infrared by spectroscopic observations with Spitzer. In contrast, the WD01 extinction curve is constructed based on interstellar dust models and constrained by photometry in the mid-infrared. Although the shape of the 8\,$\mu$m feature depends slightly on which extinction curve is employed, both extinction-corrected spectra appear more consistent with Class C because of its broad width and the absence of the 8.6\,$\mu$m feature seen in the Class A and B UIR spectra.

\subsection{Nature of the carriers}
All of the WC spectra in this paper show a broad feature around 8\,$\mm$, and the ISO/SWS WC spectra show a 6.4\,$\mm$ feature. These features are part of the UIR spectrum as discussed above. In addition to these two features, the WC spectra in Figure \ref{fig:WR125_fit} and \ref{fig:ISO_fit} exhibit excess emission above the continuum around 12--14\,$\mm$, which is especially prominent in the WR125 and WR 48a spectra. In this section, we discuss the nature of the carriers of the UIR features observed in the WC stars.

Normally the UIR features are expected to be carried by gas-phase PAH molecules and/or materials containing PAHs that emit through the process of UV-pumped IR fluorescence \citep{Allamandola85, Leger84, Tielens08}. However, in the circumstellar environment of WC stars, it may be difficult for such small molecules to survive the harsh radiation conditions as discussed in the case of dusty classical novae \citep[cf.][]{Evans94} On the other hand, bulk carbonaceous grains in the vicinity of the heating source can reach the energy equilibrium and emit the UIR features via a thermal process \citep{Dwek80,Duley11}. \citet{Endo21} showed that the time variation in the intensity of the observed 8\,$\mm$ UIR feature in the spectra of a classical nova V2361 Cyg supports their interpretation that the feature is emitted via a thermal emission process.

The Class C UIR spectra have been observed in circumstellar environments around sources such as post-AGB stars \citep{Peeters02,Diedenhoven04,Sloan07}, dusty classical novae \citep{Evans97,Evans05,Helton11,Sakon16}, and R Coronae Borealis (RCB) stars \citep{Garcia13}. The 6.2\,$\mm$ UIR feature, which is usually assigned to aromatic \ce{C-C} stretches, has been observed in these objects. The peak position of the feature shifts to longer wavelength (6.25--6.31\,$\mm$ for post-AGB stars; \citealt{Peeters02, Sloan07}, 6.31--6.41\,$\mm$ for novae; \citealt{Helton11}, and 6.24--6.41\,$\mm$ for RCB stars; \citealt{Garcia13}), which is consistent with the 6.4\,$\mm$ feature seen in the WC spectra. 

The red-shifted nature of the 6.2\,$\mu$m feature in the Class C UIR spectra could be explained by an increase in the ratio of aliphatic to aromatic content in the carriers based on the laboratory work by \citet{Pino08}. \citet{Sloan07} suggested that the carriers of the Class C UIR spectra can be a mixture of aromatic and aliphatic materials including more fragile aliphatic bonds that have not yet been exposed to intense UV radiation because aliphatic structures are expected to be broken in such harsh environments. \citet{Evans10} proposed that the UIR spectra in a dusty classical nova, DZ Cru, may be attributed to species in which aliphatic bonds predominate rather than PAH molecules. They also suggested that the carriers including fragile aliphatic structures have been protected in a dense clump and have only recently been exposed to the harsh condition. If the carriers of the UIR features around WC stars also have a high aliphatic fraction, the fragile structures could survive because they are protected in a dense region formed by wind-wind collisions and/or because the carriers are robust bulk carbonaceous grains as discussed above.

A broad 8\,$\mm$ feature has also been observed in these objects and the peak positions are 8.12--8.22\,$\mm$ for post-AGB stars \citep{Peeters02, Sloan07}, 7.77--8.17\,$\mm$ for novae \citep{Evans05, Helton11, Sakon16}, and 8.06--8.27\,$\mm$ for RCB stars \citep{Garcia13}. Recently, \citet{Endo21} indicated that the amine structure contained in the organic dust contributes to the 8\,$\mm$ feature observed in novae based on the analysis of laboratory organics called Quenched Nitrogen-included Carbonaceous Composite (QNCC), whose infrared properties can well reproduce the UIR spectra in novae. It is not certain whether it is possible to incorporate nitrogen in dust formation around WC stars because WC stars are usually nitrogen-deficient \citep{Crowther07}, while WR 48a is thought to host a WN star companion. Other than the amine structure, a sp$^{3}$ carbon defect band \citep{Gavilan17} and aromatic \ce{C=C} bonds in pyramidalized carbon \citep{Scott96,Galue14} can produce the broad 8\,$\mm$ feature.

Typical UIR features at 3.3\,$\mm$ and 11.3\,$\mm$ arising from \ce{C-H} stretching and \ce{C-H} bending modes have been detected in post-AGB stars and novae. The spectra of RCB stars by \citet{Garcia13}, which are usually hydrogen-deficient, did not show the 11.3\,$\mm$ feature except for some of the least hydrogen-deficient targets, and the spectra did not cover the 3\,$\mm$ regions. WC stars are also usually considered to be hydrogen-deficient. An emission feature at 3.28\,$\mm$ was observed in several WC stars \citep{Cohen78}. However, the feature is weaker in the spectra of WC stars with circumstellar dust shells \citep{Cohen78}, and \citet{Williams82} assigned it to {C\,{\sc iv}} ($11\rightarrow10$) transition, but not to the \ce{C-H} stretching modes of PAHs. The representative \ce{C-H} bands at 3.3 (\ce{C-H} stretch) and 11.3\,$\mm$ (\ce{C-H} bending) are not detected in any WC spectra in this paper, though the WR 125 spectrum does not cover the 3\,$\mm$ regions.

The assignment of the excess at 12--14\,$\mm$ seen in the WC spectra is not clear. Although the excess does not disappear in the range of the $A_{V}$ values considered in this paper, the wavelength region is heavily affected by the extinction correction. Some WC stars show 12.36, 12.81 and 13.12\,$\mm$ features attributable to {He\,{\sc i}}$+${He\,{\sc ii}}, [{Ne\,{\sc ii}}], and {He\,{\sc ii}} ($11\rightarrow10$), respectively \citep{Smith01}, which may contribute to part of this excess. However the FWHMs of these lines are $\lesssim 0.2 \mm$, which are much smaller than that of the observed excess.

It is known that a broad 12\,$\mm$ plateau feature is prominent in post-AGB spectra \citep{Kwok01,Joblin08}. The spectrum of the dusty classical nova V2361 Cyg at 251 days after the outburst also shows a similar plateau \citep{Helton11}. \citet{Kwok01} suggested that the 12\,$\mm$ plateau is possibly associated with C-H out-of-plane bending modes in aromatic molecules and a wide variety of alkenes connected to aromatic rings. The very weak and broad $\sim$11--15\,$\mm$ emission feature observed in H-deficient RCB spectra \citep{Garcia13} may reflect residual hydrogen in these stars. Though WC stars are also usually hydrogen-deficient, if the 12--14\,$\mm$ excess in the WC spectra includes contributions from C-H bonds, it might be possible that hydrogen is injected into the dust grains in the colliding wind region.

\section{Conclusion and Summary} \label{sec:conclusion}
We detect a broad 8\,$\mm$ feature in the WR 125 N-band low-resolution (NL) spectrum obtained with Subaru/COMICS and analyze it using two different interstellar extinction curves, WD01 and G21. In addition, we reanalyze the ISO/SWS spectra of five dusty WC stars using the same analysis method, for which the UIR features have been previously reported \citep{Cohen89, Chiar02, Marchenko17}. We show that the 8\,$\mm$ feature in the WR 125 spectrum exhibits a nearly identical shape to the 8\,$\mm$ feature in at least four of the five ISO/SWS WC stars. 

We find that the shape of the 8\,$\mm$ feature is dependent on the extinction curve used to deredden the spectra (see Figure \ref{fig:WR_comparison}). The peak positions and FWHMs of the feature in the WC spectra dereddened with the WD01 curve are at longer wavelengths and their widths are wider than those dereddened with the G21 curve. The $A_{V}$ values used for the extinction correction also affect the peak positions and FWHMs (see Table \ref{tab:WR_gaussfit}). In order to investigate the accurate shape of the feature, it is necessary to properly correct for interstellar extinction, which requires further careful study.

The features look similar despite the difference in the dust formation environments, which can be affected by the type of the companion stars or orbital properties. WR 48a may be unique in this regard, which exhibits a 8\,$\mm$ feature with the peak position at a slightly shorter wavelength. Among the dusty WC stars observed with ISO/SWS, WR 104, WR 112, and WR 118 are persistent, apparently constant dust producers, while WR 48a and WR 98a are persistent, variable dust producers.

The 8\,$\mm$ features seen in the WC spectra may be related to the Class C UIR spectra based on their broad band width and the absence of the 8.6\,$\mm$ feature seen in the Class A and Class B spectra (see Figure \ref{fig:WR_UIR}). Class C features are also seen in post-AGB stars \citep{Peeters02,Diedenhoven04,Sloan07}, dusty classical novae \citep{Evans97,Evans05,Helton11,Sakon16} and RCB stars \citep{Garcia13}. 

For future work, it is important to investigate the influence of the spectral subtype of WR and companion stars. Interestingly, WR 48a is the only dusty WC system in our sample that is thought to host a WN star companion \citep{Zhekov14}. It may be possible that difference in companion types make a slight difference in the peak position of the 8\,$\mm$ feature.

The dusty classical novae, which also show the Class C UIR features, exhibit variability in their spectra as a function of time from their outbursts \citep{Helton11}. The main target of the present study, WR 125, is an episodic/periodic dust producer, and it is expected that the infrared spectrum can vary with time after dust formation around the periastron passage. In order to better understand the nature of the UIR features of WR stars, which can possibly be a significant contributor of the band carries in the early universe, further infrared observations of other dusty WC stars and multi-epoch observations of episodic/periodic dust forming WC stars are needed.

These questions can be directly addressed with the mid-IR instrumentation on upcoming platforms such as the Tokyo Atacama Observatory \citep{Miyata10}, the \textit{James Webb Space Telescope}, and 30-m class telescopes.

\acknowledgments

We thank T. Fujiyoshi and the Subaru Observatory staff for supporting our Subaru/COMICS observations of WR 125. We also thank T. Michifuji, T. Miyata, T. Ootsubo, M. Honda, T. Fujiyoshi and T. Yamashita for the discussion on the COMICS NL spectra calibration. We thank an anonymous referee for helpful comments to improve the paper.

This research is based on data collected at Subaru Telescope, which is operated by the National Astronomical Observatory of Japan. We are honored and grateful for the opportunity of observing the Universe from Maunakea, which has the cultural, historical and natural significance in Hawaii. Based in part on observations with ISO, an ESA project with instruments funded by ESA Member States (especially the PI countries: France, Germany, the Netherlands and the United Kingdom) and with the participation of ISAS and NASA.

This publication makes use of data products from the Near-Earth Object Wide-field Infrared Survey Explorer (NEOWISE), which is a joint project of the Jet Propulsion Laboratory/California Institute of Technology and the University of Arizona. NEOWISE is funded by the National Aeronautics and Space Administration. This research is based in part on observations with AKARI, a JAXA project with the participation of ESA. When the new 2007--2008 data reported here were obtained, UKIRT was operated by the Joint Astronomy Centre on behalf of the Science and Technology Facilities Council of the U.K.

This work was supported by grants to IE from  Grant-in-Aid for JSPS Fellows (Grant No. 21J13200). RML acknowledges the Japan Aerospace Exploration Agency's International Top Young Fellowship.

\vspace{5mm}
\facilities{Subaru(COMICS), SAI 1.25-m, ISO(SWS), NEOWISE-R, AKARI(IRC), UKIRT(UIST)}

\bibliography{sample63}{}
\bibliographystyle{aasjournal}

\end{document}